\documentclass[conference]{IEEEtran}
\IEEEoverridecommandlockouts
\usepackage{amsmath, amssymb, amsthm, bm}
\usepackage{graphicx}
\usepackage{color, colortbl} 
\usepackage[dvipsnames,svgnames,x11names]{xcolor}
\makeatother
\usepackage{tabularx, boldline, multirow}
\usepackage[boxed,ruled,vlined]{algorithm2e}
\usepackage{enumerate}
\usepackage[caption=false,font=footnotesize]{subfig}
\usepackage{cite}
\usepackage{url,comment}
\usepackage[normalem]{ulem}
\usepackage[shortlabels]{enumitem}
\usepackage{pgfplots}
\usepackage{tikz}
\usepackage{mathtools}
\usepackage{accents}
\usepackage{balance}
\usepackage{acronym}
\usepackage{arydshln}
\usepackage{scalerel}

\usetikzlibrary{calc}
\makeatletter
\newcommand{\gettikzxy}[3]{%
  \tikz@scan@one@point\pgfutil@firstofone#1\relax
  \edef#2{\the\pgf@x}%
  \edef#3{\the\pgf@y}%
}

\acrodef{crb}[CRB]{Cram\'er-Rao bound}
\acrodef{awgn}[AWGN]{additive white Gaussian noise}
\acrodef{ap}[AP]{access point}
\acrodef{bs}[BS]{base station}
\acrodef{bp}[BP]{belief propagation}
\acrodef{1d}[1D]{one-dimensional}
\acrodef{3d}[3D]{three-dimensional}
\acrodef{5g}[5G]{fifth generation}
\acrodef{6g}[6G]{sixth generation}
\acrodef{ue}[UE]{user equipment}
\acrodef{iue}[IUE]{image user equipment}
\acrodef{los}[LOS]{line-of-sight}
\acrodef{aoa}[AoA]{angle-of-arrival}
\acrodefplural{aoa}[AoAs]{angle-of-arrival}
\acrodef{aod}[AoD]{angle-of-departure}
\acrodefplural{aod}[AoDs]{angle-of-departures}
\acrodef{dp}[DP]{detection probability}
\acrodefplural{dp}[DPs]{detection probabilities}
\acrodef{toa}[ToA]{time-of-arrival}
\acrodefplural{toa}[ToAs]{time-of-arrival}
\acrodef{tdoa}[TDoA]{time-difference-of-arrival}
\acrodef{ris}[RIS]{reconfigurable intelligent surface}
\acrodef{isac}[ISAC]{integrated sensing and communications}
\acrodef{nris}[NRIS]{non-RIS}
\acrodef{cp}[CP]{cyclic prefix}
\acrodef{tx}[Tx]{transmitter}
\acrodef{rx}[Rx]{receiver}
\acrodefplural{rx}[Rxs]{receivers}
\acrodef{psd}[PSD]{power spectral density}
\acrodef{crlb}[CRLB]{Cram\'er-Rao lower bounds}
\acrodefplural{crlb}[CRLBs]{Cram\'er-Rao lower bounds}
\acrodef{rss}[RSS]{received signal strength}
\acrodef{los}[LOS]{line-of-sight}
\acrodef{nlos}[NLOS]{non line-of-sight}
\acrodef{fov}[FoV]{field-of-view}
\acrodef{dft}[DFT]{discrete Fourier transform}
\acrodef{fft}[FFT]{fast Fourier transform}
\acrodef{fim}[FIM]{Fisher information matrix}
\acrodefplural{fim}[FIMs]{Fisher information matrix}
\acrodef{efim}[EFIM]{equivalent Fisher information matrix}
\acrodef{upa}[UPA]{uniform planar array}
\acrodefplural{upa}[UPAs]{uniform planar arrays}
\acrodef{peb}[PEB]{position error bound}
\acrodef{slam}[SLAM]{simultaneous localization and mapping}
\acrodef{snr}[SNR]{signal-to-noise ratio}
\acrodefplural{snr}[SNRs]{signal-to-noise ratio}
\acrodef{sre}[SRE]{smart radio environment}
\acrodefplural{sre}[SRE]{smart radio environments}
\acrodef{mpmb}[MPMB]{marginal Poisson multi-Bernoulli}
\acrodef{pmb}[PMB]{Poisson multi-Bernoulli}
\acrodef{mimo}[MIMO]{multiple-input  multiple-output}
\acrodef{rfs}[RFS]{random finite set}
\acrodef{rfid}[RFID]{radio-frequency identification}
\acrodef{siso}[SISO]{single-input single-output}
\acrodef{miso}[MISO]{multiple-input single-output}
\acrodef{ici}[ICI]{inter-carrier interference}
\acrodef{iid}[iid]{independent and identically distributed}
\acrodef{ml}[ML]{maximum likelihood}
\acrodef{pdf}[PDF]{probability density function}
\acrodef{cdf}[CDF]{cumulative distribution function}
\acrodef{ccdf}[CCDF]{complementary cumulative distribution function}
\acrodef{ofdm}[OFDM]{orthogonal frequency-division multiplexing}
\acrodef{qos}[QoS]{Quality of Service}
\acrodef{sp}[SP]{scattering point}
\acrodefplural{sp}[SPs]{scattering points}
\acrodef{nsp}[NSP]{near scattering point}
\acrodefplural{nsp}[NSPs]{near scattering points}
\acrodef{ula}[ULA]{uniform linear array}
\acrodef{va}[VA]{virtual anchor}
\acrodefplural{va}[VAs]{virtual anchors}
\acrodef{ls}[LS]{large surface}
\acrodefplural{ls}[LSs]{large surfaces}
\acrodef{rp}[RP]{reflection point}
\acrodefplural{ls}[RPs]{reflection point}
\acrodef{eb}[EB]{error bound}
\acrodefplural{eb}[EBs]{error bounds}
\acrodef{peb}[PEB]{position error bound}
\acrodefplural{peb}[PEBs]{position error bound}
\acrodef{heb}[HEB]{heading error bound}
\acrodefplural{heb}[HEBs]{heading error bound}
\acrodef{seb}[SEB]{speed error bound}
\acrodefplural{seb}[SEBs]{speed error bound}
\acrodef{mae}[MAE]{mean absolute error}
\acrodefplural{mae}[MAEs]{mean absolute error}
\acrodef{ab}[AB]{arbitrary beam}
\acrodef{cb}[CB]{conventional beam}
\acrodef{gospa}[GOSPA]{generalized optimal subpattern assignment}
\acrodef{mae}[MAE]{mean absolute error}
\acrodef{ppp}[PPP]{Poisson point process}
\acrodef{gci}[GCI]{generalized covariance intersection}

\pgfplotsset{compat=1.18}
\begin{document}
\bstctlcite{IEEEexample:BSTcontrol}

\title{RIS-Aided Monostatic Sensing and Object Detection with Single and Double Bounce Multipath}

\author{Hyowon Kim\IEEEauthorrefmark{1}, Alessio Fascista\IEEEauthorrefmark{2}, Hui Chen\IEEEauthorrefmark{1}, Yu Ge\IEEEauthorrefmark{1},\\ George C. Alexandropoulos\IEEEauthorrefmark{3}, 
Gonzalo Seco-Granados\IEEEauthorrefmark{4}, 
and Henk Wymeersch\IEEEauthorrefmark{1}\\
\IEEEauthorrefmark{1}Chalmers University of Technology, Gothenburg, Sweden, 
\IEEEauthorrefmark{2}Universit{\`a} del Salento, Italy\\
\IEEEauthorrefmark{3}National and Kapodistrian University of Athens, Greece, 
\IEEEauthorrefmark{4}Universitat Autonoma de Barcelona, Spain 
\thanks{This work was supported, in part, by the EU H2020 RISE-6G project under grant 101017011, by the Chalmers Area of Advance Transport 6G-Cities project, by Basic Science Research Program through the NRF of Korea (2022R1A6A3A03068510), and by MSCA-IF grant 101065422 (6G-ISLAC).}
}

\maketitle

\begin{abstract}
    We propose a framework for monostatic sensing by a user equipment (UE), aided by a reconfigurable intelligent surface~(RIS) in environments with single- and double-bounce signal propagation. 
    We design appropriate UE-side precoding and combining, to facilitate signal separation. We derive the adaptive detection probabilities of the resolvable signals, based on the geometric channel parameters of the links.
    Then, we estimate the passive objects using both the double-bounce signals via passive RIS (i.e., RIS-sensing) and the single-bounce multipath direct to the objects (i.e., non-RIS-sensing), based on a mapping filter. 
    Finally, we provide numerical results to demonstrate that effective sensing can be achieved through the proposed framework.
\end{abstract}

\section{Introduction}
    \Ac{isac} is 
    expected to be a key functionality
    in \ac{6g} communications, enabling a variety of applications~\cite{An_ISAC_Survey2022}.
    \Acp{ris} facilitate \ac{isac} thanks to the enhanced coverage, obtained by reflecting the received signal power, or to the creation of a  controllable wireless propagation environment by proper design of the phase profiles \cite{chepuri2022integrated,Hyowon_RISSLAM_TWC2022},
    representing thus one of the \ac{6g} enablers~\cite{Emil_RIS_SPM2022}.

\begin{figure}
    \centering
    {\includegraphics[width=0.85\columnwidth]{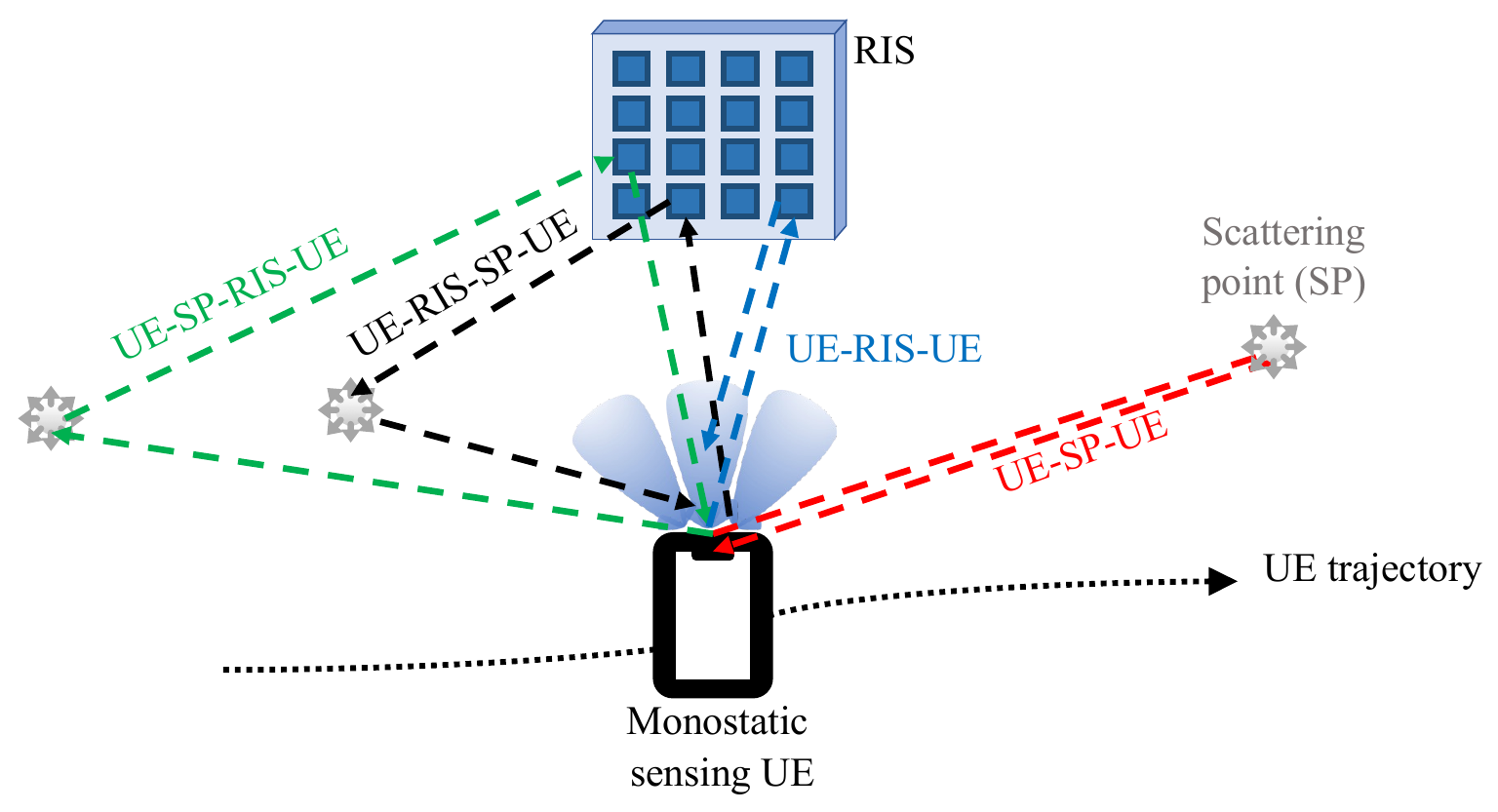}}
     \vspace{-0.2cm}
    \caption{Illustration of the considered sensing  scenario where a single monostatic UE maps the environment (described by scattering points) with the support of an RIS.}
     \vspace{-.6cm}
    \label{Fig:scenario}
\end{figure}

    Monostatic sensing with RIS is relatively under-explored. Relevant works in this direction include 
     \cite{Zhang_RISSL_Proc2022,Ziang_RISSLAM_TWC2022,aubry2021reconfigurable,buzzi2021radar,buzzi2022foundations,wang2021joint,sankar2021joint,alexandropoulos2021hybrid}. 
    The authors of~\cite{Zhang_RISSL_Proc2022} introduce case studies of \ac{ris}-enabled sensing and localization, including the double-bounce signal scenario, where the signal reflected by the RIS can impinge on a \ac{sp} before being received back at the UE (denoted by UE-RIS-SP-UE). In~\cite{Ziang_RISSLAM_TWC2022}, several \acp{ris} are regarded as controllable passive objects with a priori unknown location. Paths of the form UE-SP-UE and UE-RIS-UE are considered to map the environment and localize the UE. In \cite{aubry2021reconfigurable}, an RIS is used to overcome \ac{los} blockage in radar sensing. Radar performance is further studied in \cite{buzzi2021radar,buzzi2022foundations}, focusing on a single-SP scenario, which simplifies the problem significantly. Studies focused on \ac{isac} include 
    \cite{wang2021joint,sankar2021joint}, where in \cite{wang2021joint} an \ac{ris} is used to reduce multi-user interference at the \acp{ue} due to the joint radar and communication signal sent by a \ac{bs}, while \cite{sankar2021joint} considers allocating separated  \ac{ris} elements between sensing and communications. Finally \cite{alexandropoulos2021hybrid} goes even further and considers a hybrid \ac{ris} that can actively sense the environment. 
    %
    %
    Despite these studies on RIS-aided  sensing, there are still several unsolved problems in the monostatic regime (see Fig.~\ref{Fig:scenario}): how to separate the different single and double-bounce signals corresponding to the several SPs; how to design UE precoders and combiners to enable tractable processing; how to fuse information coming from single-bounce and double-bounce signals associated with a single SP; and how much RIS can help when a priori information about the SPs is unavailable.


    
    In this paper, we propose a framework of \ac{ris}-aided monostatic range-angle sensing to estimate the locations of several passive objects (i.e., \acp{sp}). The contributions are as follows: (i) we derive the signal model that encompasses all single- and double-bounce paths via the RIS; (ii) to enable separation of the different paths, we propose a suitably-designed \ac{ue}-side precoding and combining scheme; (iii) we derive analytical expressions for the \acp{dp} of the objects, based on the separated observations; (iv) finally, we fuse the different observations and map the \acp{sp} as the \ac{ue} explores the environment. This fusion is based on two state-of-the-art \ac{pmb} filters~\cite{Garcia-Fernandez2018,Hyowon_MPMB_TVT2022}, 
    one with the double-bounce signals via passive \ac{ris}; and the other with the single-bounce signals direct to \acp{sp}. Sensing results are finally merged into one map by the \ac{gci} fusion method~\cite{Batistelli_GCI_JSTSP2013}. Numerical results reveal that, under the considered UE precoding and combining and random RIS configurations, the single-bounce path provides the most information about the SPs, followed by the path UE-RIS-SP-UE, while the path UE-SP-RIS-UE is  less informative. 


\section{System and Signal Models}
    In this section, we introduce proper models for monostatic sensing and object detection, aided by a single RIS.

\subsection{System Setup}
    Consider the generic scenario adapted from \cite[Fig.~1 (b) and (d)]{buzzi2022foundations}
    in Fig.~\ref{Fig:scenario}, where a full-duplex \ac{ue} transmits a signal using an antenna array and receives the backscattered signal from both passive objects (\acp{sp}) and an \ac{ris}. Under this scenario, there are at least four different types of paths:  two single-bounce paths, which are the path via the RIS, namely the path, UE-RIS-UE (shown in blue), and the conventional radar paths, UE-SP-UE (shown in red). There are also two double-bounce paths per SP, namely the path, UE-RIS-SP-UE (shown in black), and the path, UE-SP-RIS-UE (shown in green). Higher-order bounces are  ignored, as they are much weaker. Hence, each SP can be observed up to 3 times, depending on the corresponding end-to-end \acp{snr}.
    Here, the signals for i) UE-RIS-UE, ii) UE-RIS-SP-UE, iii) UE-SP-RIS-UE can be controlled by the \ac{ris} while those in iv) UE-SP-UE cannot.
    The full-duplex \ac{ue} and the \ac{ris} are equipped with \acp{upa}, and their array sizes are respectively $N_\mathrm{U} = N_\mathrm{U}^\mathrm{az}\times N_\mathrm{U}^\mathrm{el}$ and $N_\mathrm{R} = N_\mathrm{R}^\mathrm{az}\times N_\mathrm{R}^\mathrm{el}$.
    
    We denote the \ac{ue} state at epoch\footnote{Epochs refer to slow time (e.g, s-level) and are indexed with $k$, while transmissions refer to fast time (e.g., $\mu$s-level) and are indexed with $t$.} $k$ by $\mathbf{s}_{k}=[\mathbf{x}_{\mathrm{U},k}^\top,\alpha_{\mathrm{U},k},v_{\mathrm{U},k}]^\top$, where the elements are the location, heading, and speed, respectively.
    The \ac{ris} location is denoted by $\mathbf{x}_\mathrm{R}$, and the $l$-th \ac{sp} location is denoted by $\mathbf{x}^l$.
    To handle the unknown number of \acp{sp} and their locations that specify the propagation environment, we model the \acp{sp} by a \ac{rfs} $\mathcal{X}=\{\mathbf{x}^1,\dots,\mathbf{x}^n\}$, with the set density $f(\mathcal{X})$~\cite{mahler_book_2007}.
    We assume the \ac{ue} state~(location, heading, and speed) and \ac{ris} location are known, to focus on the sensing performance.
    
\subsection{Signal and Channel Models}
    We adopt a deterministic channel model that considers only large-scale fading for all
    resolvable paths. The received \ac{ofdm} signal of the $s$-th subcarrier at the $t$-th transmission of time epoch $k$ is modeled as\footnote{We consider all $L$ SPs to be present at all time, though they may not all be detectable at each epoch, $k$. }
\begin{align}
    \mathbf{y}_{k,t,s} 
    \triangleq &
    \bigg( \sum_{l=0}^{L}
    \underbrace{\alpha_{k,l}\nu_{t}(\boldsymbol{\phi}_{k,l})
    \mathbf{a}_\mathrm{U}(\boldsymbol{\theta}_{k,l}) \mathbf{a}_\mathrm{U}^\top(\boldsymbol{\theta}_{k,0}) 
    d_s(\tau_{k,l})
    }_{\text{i) UE-RIS-UE}~(l=0),~\text{ii) UE-RIS-SPs-UE}~(l\neq 0)} \label{eq:signal}\\ 
    & + \sum_{l=1}^{L}
    \underbrace{\alpha_{k,l}\nu_{t}(\boldsymbol{\phi}_{k,l})
    \mathbf{a}_\mathrm{U}(\boldsymbol{\theta}_{k,0}) \mathbf{a}_\mathrm{U}^\top(\boldsymbol{\theta}_{k,l}) 
    d_s(\tau_{k,l})
    }_{{\text{iv) UE-SPs-RIS-UE~paths}}} \notag \\ 
    & + \sum_{l=1}^L
    \underbrace{\beta_{k,l}
    \mathbf{a}_\mathrm{U}(\boldsymbol{\theta}_{k,l}) \mathbf{a}_\mathrm{U}^\top(\boldsymbol{\theta}_{k,l}) 
    d_s(\bar{\tau}_{k,l})
    }_{\text{iii) UE-SPs-UE~paths}} \bigg) {\mathbf{f}}_{k,t} 
    + \mathbf{n}_{k,t,s}.\notag
\end{align}
    The parameters $\alpha_{k,l}$ and $\beta_{k,l}$ are respectively the complex path gains of the controlled and uncontrolled signals; $\mathbf{a}_\mathrm{R}(\cdot)$ and $\mathbf{a}_\mathrm{U}(\cdot)$ are respectively the array vectors~\cite[eqs.~(13)--(15)]{Kamran_RISMobility_JSTSP2022} of the \ac{ris} and \ac{ue} with  $\boldsymbol{\phi}_{k,l}=[\phi_{k,l}^\mathrm{az},\phi_{k,l}^\mathrm{el}]^\top$ denoting the azimuth and elevation of the \ac{aoa} and \ac{aod}\footnote{In this monostatic scenario, the \ac{aoa} is identical to the \ac{aod}.} at the \ac{ris} and $\boldsymbol{\theta}_{k,l}=[\theta_{k,l}^\mathrm{az},\theta_{k,l}^\mathrm{el}]^\top$ the \ac{aoa} and \ac{aod} at the \ac{ue}; $d_s(\tau) \triangleq e^{-j2\pi (s-1) \tau \Delta_f }$ is the phase shift linked to the \ac{toa} with $\tau_{k,l}$ and $\bar{\tau}_{k,l}$ denoting the \acp{toa} for the controlled and uncontrolled signals; $\Delta_f$ denotes the subcarrier spacing; ${\mathbf{f}}_{k,{t}}$ denotes the precoder with $\Vert {\mathbf{f}}_{k,{t}}\Vert^2=1$; and $\mathbf{n}_{k,t,s} \sim \mathcal{CN}(\boldsymbol{0}_{\mathrm{UE}_N},N_0\mathbf{I}_{N_\mathrm{U}})$ denotes the complex Gaussian noise.
    Finally, we denote the \ac{ris} phase profile $\boldsymbol{\Omega}_{\mathrm{R},k,t} \triangleq \mathrm{diag}(\boldsymbol{\omega}_{k,t})$, where $\boldsymbol{\omega}_{k,t}\triangleq [\omega_{k,t}^1,\dots,\omega_{k,t}^{N_\mathrm{R}}]^\top$, so that  $\nu_{t}(\boldsymbol{\phi}_{k,l}) \triangleq  
    \mathbf{a}_\mathrm{R}^\top(\boldsymbol{\phi}_{k,l})
    \boldsymbol{\Omega}_{\mathrm{R},k,t} 
    \mathbf{a}_\mathrm{R}(\boldsymbol{\phi}_{k,0})$.
    The channel parameters are defined in Appendix~\ref{app:ChPara}. In the following, we assume that $T /\Delta_f$ is sufficiently small so that Doppler effects can be considered negligible.

\subsection{Precoders and RIS Phase Profiles}
    The precoders and the \ac{ris} phase profiles follow a specific time sequence~\cite{Kamran_RISloc_ICC2022,Kamran_RISMobility_JSTSP2022} for orthogonal 
    design (that indicates $\sum_t\bm{\omega}_{t,k}=0$): $\tilde{\bm{\omega}}_{\tilde{t},k} \in \mathbb{C}^{N_\mathrm{R} \times 1}$ for $\tilde{t}=1,\dots,T/2$, and $\bm{\omega}_{2\tilde{t}-1,k} \triangleq \tilde{\bm{\omega}}_{\tilde{t},k}$,  $\bm{\omega}_{2\tilde{t},k} \triangleq -\tilde{\bm{\omega}}_{\tilde{t},k}$,
    and $\mathbf{f}_{2\tilde{t}-1,k}\triangleq\mathbf{f}_{2\tilde{t},k} \triangleq \tilde{\mathbf{f}}_{\tilde{t},k}$.
    The $n$-th element of \ac{ris} phase profile is denoted by $[\tilde{w}_{\tilde{t},k}]^n = e^{j\phi_{\tilde{t},k}^n}$, where $\phi_{\tilde{t},k}^n$ is the \ac{ris} phase.

\section{Signal Separation}
    In this section, we propose an approach for separating the different contributions in the received signal  \eqref{eq:signal}. 

\subsection{RIS and Non-RIS Signals}
    By leveraging the orthogonal 
    \ac{ris} phase design, we divide the received signals into
    the controlled~(i--iii) and uncontrolled (iv) signals as follows, for $\tilde{t}=1,\dots,T/2$
\begin{align}
   &  \tilde{\mathbf{y}}_{k,\tilde{t},s}^\mathrm{R} \triangleq  \, \frac{1}{2}(\mathbf{y}_{k,2t,s} - \mathbf{y}_{k,2t-1,s}) \\
    & = \, \sum_{l=0}^{L}
    \alpha_{k,l}\nu_{\tilde{t}}(\boldsymbol{\phi}_{k,l})
    \mathbf{a}_\mathrm{U}(\boldsymbol{\theta}_{k,l}) \mathbf{a}_\mathrm{U}^\top(\boldsymbol{\theta}_{k,0}) 
    d_s(\tau_{k,l})
    \tilde{\mathbf{f}}_{k,\tilde{t}} \label{eq:NRISpaths} \\
    &+ \sum_{l=1}^{L}
    \alpha_{k,l}\nu_{\tilde{t}}(\boldsymbol{\phi}_{k,l})
    \mathbf{a}_\mathrm{U}(\boldsymbol{\theta}_{k,0}) \mathbf{a}_\mathrm{U}^\top(\boldsymbol{\theta}_{k,l}) 
    d_s(\tau_{k,l})
    \tilde{\mathbf{f}}_{k,\tilde{t}}+\tilde{\mathbf{n}}_{k,\tilde{t},s}^\mathrm{R},\notag
\end{align}    
and
\begin{align}
    \tilde{\mathbf{y}}_{k,\tilde{t},s}^\mathrm{N} \triangleq & \, \frac{1}{2}(\mathbf{y}_{k,2t,s} + \mathbf{y}_{k,2t-1,s})\\
    =& \, \sum_{l=1}^L
    \beta_{k,l}\mathbf{a}_\mathrm{U}(\boldsymbol{\theta}_{k,l}) \mathbf{a}_\mathrm{U}^\top(\boldsymbol{\theta}_{k,l}) 
    d_s(\bar{\tau}_{k,l}) \tilde{\mathbf{f}}_{k,\tilde{t}}+\tilde{\mathbf{n}}_{k,\tilde{t},s}^\mathrm{N},\label{eq:NRIS}
\end{align}
    where 
    $\tilde{\mathbf{n}}_{k,\tilde{t},s}^\mathrm{R}$ and $\tilde{\mathbf{n}}_{k,\tilde{t},s}^\mathrm{N}$ are independent complex Gaussian noise  contributions, distributed as
    $\mathcal{CN}(\boldsymbol{0}_{N_\mathrm{U}},N_0\mathbf{I}_{N_\mathrm{U}}/2)$.
    We assume that the RIS signal is always visible, and  $\boldsymbol{\phi}_{k,0}$ and $\boldsymbol{\theta}_{k,0}$ are known, due to the knowledge of the UE state. 

\subsection{Separation of the RIS Signals}
In \eqref{eq:NRISpaths}, the path to and from the RIS appear together, so that without suitable processing, up to $2L+1$ path will be present. To avoid this, we propose a method to separate the UE-SP-RIS-UE paths (second term in \eqref{eq:NRISpaths}) from the UE-RIS-SP-UE paths (first term in \eqref{eq:NRISpaths}), by designing the precoder and combiner at the UE, inspired by the approach from \cite{palmucci2022ris}.
In particular, we divide up the $T/2$ available transmissions into  $T_1$ \emph{transmissions towards the RIS}, with $\tilde{\mathbf{f}}_{k,\tilde{t}}= \mathbf{a}^*_\mathrm{U}(\boldsymbol{\theta}_{k,0})/\Vert \mathbf{a}_\mathrm{U}(\boldsymbol{\theta})\Vert $ and $T_2=T/2-T_1$ transmissions with a \emph{null towards the RIS}, i.e., with $\tilde{\mathbf{f}}_{k,\tilde{t}}^{\mathrm{H}} \mathbf{a}_\mathrm{U}(\boldsymbol{\theta}_{k,0})=0$.\footnote{Such precoders can be designed through orthogonal projection onto the null space of $\mathbf{a}_\mathrm{U}(\boldsymbol{\theta}_{k,0})$.} For each transmission, we use the (invertible and thus lossless) combiner $\mathbf{W}_k = [\mathbf{a}_\mathrm{U}(\boldsymbol{\theta}_{k,0})/\Vert \mathbf{a}_\mathrm{U}(\boldsymbol{\theta})\Vert, \mathbf{W}_{k,\perp}]$ such that $\mathbf{W}_k^\mathrm{H}\mathbf{W}_k=\mathbf{I}_{ N_\mathrm{U}}$ and $\mathbf{W}_{k,\perp}^{\mathrm{H}} 
    \mathbf{a}_\mathrm{U}(\boldsymbol{\theta}_{k,0})=\boldsymbol{0}_{N_\mathrm{U}-1}$.


\subsubsection{Observation during \texorpdfstring{$T_1$}{TEXT}~Transmissions toward RIS}
\label{sec:URIS}
    During the transmissions when $\tilde{\mathbf{f}}_{k,\tilde{t}}= \mathbf{a}^*_\mathrm{U}(\boldsymbol{\theta}_{k,0})/\Vert \mathbf{a}_\mathrm{U}(\boldsymbol{\theta}\Vert $, the output of the combiner will be $\mathbf{W}^{\mathrm{H}}_k\tilde{\mathbf{y}}_{k,\tilde{t},s}^\mathrm{R}$. 
    To remove the unwanted UE-RIS-UE path, we discard the first entry\footnote{Note that discarding the first entry in $\mathbf{W}^{\mathrm{H}}_k\tilde{\mathbf{y}}_{k,\tilde{t},s}^\mathrm{R}$ leads to a loss of information for SPs that are on the line between the UE and the RIS.} and denote the remainder by $\mathbf{\bar{y}}_{k,\tilde{t},s}^\mathrm{D} \in \mathbb{C}^{N_\mathrm{U} \times 1}$ (`D' is used for directional), which is expressed as 
\begin{align}
  \mathbf{\bar{y}}_{k,\tilde{t},s}^\mathrm{D}  & =\mathbf{W}^{\mathrm{H}}_{k,\perp}\tilde{\mathbf{y}}_{k,\tilde{t},s}^\mathrm{R} =   \mathbf{W}^{\mathrm{H}}_{k,\perp}
    \tilde{\mathbf{n}}_{k,\tilde{t},s}^\mathrm{R} \label{eq:reURIS}\\
 & +\sqrt{N_\mathrm{U}}\sum_{l=0}^{L}
    \alpha_{k,l}\nu_{\tilde{t}}(\boldsymbol{\phi}_{k,l})
    \mathbf{W}^{\mathrm{H}}_{k,\perp}\mathbf{a}_\mathrm{U}(\boldsymbol{\theta}_{k,l}) 
    d_s(\tau_{k,l})
     \notag 
\end{align}
since $\mathbf{a}_\mathrm{U}^\top(\boldsymbol{\theta}_{k,0}) 
    \tilde{\mathbf{f}}_{k,\tilde{t}}=\Vert \mathbf{a}_\mathrm{U}(\boldsymbol{\theta}_{k,0})\Vert=\sqrt{N_\mathrm{U}}$.

\subsubsection{Observation during \texorpdfstring{$T_2$}{TEXT}~Transmissions with null to RIS}
\label{sec:SRIS}
During the remaining $T_2$ transmissions, the (arbitrary)  precoders with null towards the RIS ensures that the first term in \eqref{eq:NRISpaths} is cancelled. In the observation after combining $\mathbf{W}^{\mathrm{H}}_k\tilde{\mathbf{y}}_{k,\tilde{t},s}^\mathrm{R}$, only the first entry contains information, since the remaining part $\mathbf{W}^{\mathrm{H}}_{k,\perp}\tilde{\mathbf{y}}_{k,\tilde{t},s}^\mathrm{R}$ only contains noise. Hence, the useful observation is $\bar{{y}}_{k,\tilde{t},s}^{\mathrm{O}}\in \mathbb{C}$ (`O' is used for orthogonal), with 
\begin{align}
    \bar{{y}}_{k,\tilde{t},s}^{\mathrm{O}} & = \frac{\mathbf{a}_\mathrm{U}^{\mathrm{H}}(\boldsymbol{\theta}_{k,0})}{\Vert \mathbf{a}_\mathrm{U}\Vert }\tilde{\mathbf{y}}_{k,\tilde{t},s}^{\mathrm{R}}  =    \frac{\mathbf{a}_\mathrm{U}^{\mathrm{H}}(\boldsymbol{\theta}_{k,0})}{\Vert \mathbf{a}_\mathrm{U}\Vert }  \tilde{\mathbf{n}}_{k,\tilde{t},s}^\mathrm{R} \label{eq:reSRIS} \\
    &+   \sqrt{N_{\mathrm{U}}}\sum_{l=1}^{L}
    \alpha_{k,l}
    \nu_{\tilde{t}}(\boldsymbol{\phi}_{k,l}) 
    \mathbf{a}_\mathrm{U}^\top(\boldsymbol{\theta}_{k,l}) 
    d_s(\tau_{k,l})\tilde{\mathbf{f}}_{k,\tilde{t}}. \notag
\end{align}

    In summary, we  obtain the three types of refined signals corresponding to ii)--iv): 
    ii) UE-RIS-SPs-UE to $\bar{\mathbf{y}}_{k,\tilde{t},s}^\mathrm{D}$ of~\eqref{eq:reURIS}, during $T_1$ transmissions ($\tilde{t}=1,\dots,T_1$); iii) UE-SPs-RIS-UE to ${{y}}_{t,\bar{t},s}^\mathrm{O}$ of~\eqref{eq:reSRIS}, during $T_2=T/2-T_1$ transmissions ($\tilde{t}=T_1+1,\dots,T/2$); and iv) UE-SPs-UE to $\tilde{\mathbf{y}}_{k,\tilde{t},s}^\mathrm{N}$ of~\eqref{eq:NRIS}, during $T/2$ transmissions ($\tilde{t}=1,\dots,T/2$).

    
\section{Detection Probability}
\label{sec:PD}
    We compute the \acp{dp}, $p_\mathrm{U}^
    \mathrm{D}(\mathbf{x}^l,\mathbf{s}_k)$, $p_\mathrm{U}^
    \mathrm{O}(\mathbf{x}^l,\mathbf{s}_k)$, and $p_\mathrm{U}^
    \mathrm{N}(\mathbf{x}^l,\mathbf{s}_k)$, for all paths.
    Following \cite{wymeersch2020adaptive}, we will focus on a single path at each signal and omit the time index $k$ for  notational simplicity.
    
\subsection{Hypothetical Observation}
    Paths in the separated signals ii)--iv)
    are expressed as
\begin{align}
    \mathbf{\bar{y}}_{l,\tilde{t},s}^\mathrm{D} \triangleq  
    & \, 
    \alpha_{l}\sqrt{N_\mathrm{U}} \nu_{\tilde{t}}(\boldsymbol{\phi}_{l})
    \mathbf{W}_\perp^\mathrm{H}
    \mathbf{a}_\mathrm{U}(\boldsymbol{\theta}_{l})  d_s(\tau_{l})
     + 
     \mathbf{\bar{n}}_{l,\tilde{t},s}^\mathrm{D}
   ,\label{eq:refinedU}\\
    \bar{{y}}_{l,\tilde{t},s}^{\mathrm{O}} \triangleq & 
    \,   \alpha_{l}\sqrt{N_\mathrm{U}}
    \nu_{\tilde{t}}(\boldsymbol{\phi}_{l})
    \mathbf{a}_\mathrm{U}^\top(\boldsymbol{\theta}_{l}) 
    \tilde{\mathbf{f}}_{\tilde{t}}d_s(\tau_{l}) 
 +\bar{{n}}_{l,\tilde{t},s}^{\mathrm{O}} ,
    \label{eq:refinedS}\\
    \tilde{\mathbf{y}}_{l,\tilde{t},s}^\mathrm{N} \triangleq&\, \beta_{l}
    \mathbf{a}_\mathrm{U}(\boldsymbol{\theta}_{l}) \mathbf{a}_\mathrm{U}^\top(\boldsymbol{\theta}_{l}) 
    \tilde{\mathbf{f}}_{\tilde{t}}d_s(\bar{\tau}_{l})+ \tilde{\mathbf{n}}_{\tilde{t},s}^\mathrm{N}.\label{eq:refinedN}
\end{align}
where $\mathbf{\bar{n}}_{l,\tilde{t},s}^\mathrm{D}= \mathbf{W}_\perp^\mathrm{H}
    \tilde{\mathbf{n}}_{\tilde{t},s}$ with $\mathbb{E}\{\mathbf{\bar{n}}_{l,\tilde{t},s}^\mathrm{D} (\mathbf{\bar{n}}_{l,\tilde{t},s}^\mathrm{D})^{\mathrm{H}}\}=N_0 \mathbf{I}_{N_{\mathrm{U}}-1}$, $\bar{{n}}_{l,\tilde{t},s}^{\mathrm{O}}=  
    \frac{\mathbf{a}_\mathrm{U}^{\mathrm{H}}(\boldsymbol{\theta}_{0})}{\Vert \mathbf{a}_\mathrm{U}\Vert } 
    \tilde{\mathbf{n}}_{\tilde{t},s}$ with $\mathbb{E}\{\bar{{n}}_{l,\tilde{t},s}^{\mathrm{O}} (\bar{{n}}_{l,\tilde{t},s}^{\mathrm{O}})^{\mathrm{H}}\}=N_0$. 
    We derive the detection probability related to $\mathbf{\bar{y}}_{l,\tilde{t},s}^\mathrm{D}$, while the other observations can be treated similarly. Let us define
\begin{align}
    \mathbf{p}_{l,\tilde{t},s}^\mathrm{D} \triangleq &
     \sqrt{N_\mathrm{U}} \nu_{\tilde{t}}(\boldsymbol{\phi}_{l})
    \mathbf{W}_\perp^\mathrm{H}
    \mathbf{a}_\mathrm{U}(\boldsymbol{\theta}_{l})  d_s(\tau_{l})
\end{align}
    and then introduce compressed observations for the signals ii)--iv),
    computed by coherent combining over subcarriers and
    transmissions $\tilde{t}=1,\dots,T_1$ for ii), $\tilde{t}=T_1+1,\dots,T/2$ for iii), and $\tilde{t}=1,\dots,T/2$ for iv) as follows:
    $\rho_l^\mathrm{D} 
    \triangleq \sum_{\tilde{t},s} (\mathbf{p}_{l,\tilde{t},s}^\mathrm{D})^\mathrm{H} \bar{\mathbf{y}}_{l,\tilde{t},s}^\mathrm{D}$ (and similarly
    $\rho_{l}^{\mathrm{O}} \triangleq \sum_{\tilde{t},s} ({p}_{l,\tilde{t},s}^\mathrm{O})^\mathrm{H} \bar{{y}}_{l,\tilde{t},s}^{\mathrm{O}}$, and 
    $\rho_{l}^\mathrm{N}
    \triangleq \sum_{\tilde{t},s} (\mathbf{p}_{l,\tilde{t},s}^\mathrm{N})^\mathrm{H} \tilde{\mathbf{y}}_{l,\tilde{t},s}^\mathrm{N}$).
    The observations are represented as $ \rho_{l}^\mathrm{D} =\sum_{\tilde{t},s} \alpha_{l} 
    \lVert \mathbf{p}_{l,\tilde{t},s}^\mathrm{D} \rVert^2 
    + w_{l}^\mathrm{D},$
    where the noise terms are defined as $w_{l}^\mathrm{D} \triangleq  \sum_{\tilde{t},s} (\mathbf{p}_{l,\tilde{t},s}^\mathrm{D})^\mathrm{H}
     \mathbf{\bar{n}}_{l,\tilde{t},s}^\mathrm{D}$. 
    We obtain new observations as follows: $\breve{y}_{l}^\mathrm{D}  \triangleq 
    \rho_{l}^\mathrm{D} / \sqrt{\mathbb{E}[ \lvert w_{l}^\mathrm{D}\rvert^2]/2}$, 
    where the expectation is computed as $ \mathbb{E}[ \lvert w_{l}^\mathrm{D}\rvert^2] =  \tilde{P}_{l}^\mathrm{D}N_0/2$,  
    in which $\tilde{P}_{l}^\mathrm{D} \triangleq \sum_{\tilde{t},s}\lVert \mathbf{p}_{l,\tilde{t},s}^\mathrm{D}
    \rVert^2$.

\subsection{Detection Probabilities with Hypothetical Statistics}
    Now, we consider hypothetical statistics for 
    signals ii)--iv) denoted as $\lvert \breve{{y}}_{l}^{\mathrm{D}} \rvert^2$,
    $\lvert \breve{{y}}_{l}^{\mathrm{O}} \rvert^2$, and $\lvert \breve{{y}}_{l}^{\mathrm{N}} \rvert^2$,
    which follow non-central chi-squared distribution with non-centrality parameter $\lambda_{l}^{\mathrm{D}}  = 4\alpha_{l}^2
    \tilde{P}_{l}^{\mathrm{D}}/N_0$ (and similarly $\lambda_{l}^{\mathrm{O}}$ and $\lambda_{l}^{\mathrm{N}} $).
    %
%
%
    %
%
    Finally, the \ac{dp} for the $l$-th path of UE-RIS-SP-UE signal 
    is computed as \cite[eq.(13)]{buzzi2022foundations}
\begin{align}
     p_\mathrm{U}^\mathrm{D}(\mathbf{x}^l,\mathbf{s}) &= f(\lvert \breve{y}_{l}^\mathrm{D} \rvert ^2 > \gamma ) 
    = Q_1\Big( \sqrt{
    \lambda_l^\mathrm{D}
    }
    ,\sqrt{\gamma}\Big),\label{eq:ADP}
\end{align}
    where $Q_1(\cdot,\cdot)$ denotes the Marcum Q-function, and $\gamma = -2\log p_\mathrm{FA}$, in which $p_\mathrm{FA}$ is the false alarm probability. Similarly, $p_\mathrm{U}^\mathrm{O}(\mathbf{x}^l,\mathbf{s})$ and $p_\mathrm{U}^\mathrm{N}(\mathbf{x}^l,\mathbf{s})$ can be computed.



\section{Poisson Multi-Bernoulli Filtering for Passive Object Sensing}

    We first describe the measurements from the separated signals.
    Since each SP can give rise to up to 3 paths and thus 3 measurements, two problems occur: a \emph{data association} problem concerning  which UE-SP-UE, UE-RIS-SP-UE, and UE-SP-RIS-UE paths are related to the same SP, and a \emph{fusion} problem regarding how to combine the associated measurements. 
    To address these problems,
    we associate and fuse all the double-bounce measurements using ellipsoidal gating.
    Then, using the measurements, we run two independent \ac{pmb} filters: one with single- and the other with double-bounce measurements.
    Finally, we perform periodic fusion.
    When possible, we will omit the time index $k$. 
 
    \subsection{Measurements}
    By applying a channel estimation routine at each time $k$ on the  signals $\bar{\mathbf{y}}_{k,\tilde{t},s}^\mathrm{D}$, $\bar{{y}}_{k,\tilde{t},s}^\mathrm{O}$, and $\tilde{\mathbf{y}}_{k,\tilde{t},s}^\mathrm{N}$, we respectively obtain channel parameter sets
    $\mathcal{Z}_k^\mathrm{D}\triangleq\{\mathbf{z}_{k,1}^{\mathrm{D}},\dots,\mathbf{z}_{k,J_k^\mathrm{D}}^{\mathrm{D}}\}$,
    $\mathcal{Z}_k^\mathrm{O}\triangleq\{\mathbf{z}_{k,1}^{\mathrm{O}},\dots,\mathbf{z}_{k,J_k^\mathrm{O}}^{\mathrm{O}}\}$, and
    $\mathcal{Z}_k^\mathrm{N}\triangleq\{\mathbf{z}_{k,1}^{\mathrm{N}},\dots,\mathbf{z}_{k,J_k^\mathrm{N}}^{\mathrm{N}}\}$, where $J_k^\mathrm{D}$, $J_k^\mathrm{O}$, and $J_k^\mathrm{N}$ are the number of detected paths (based on the computed detection probabilities from Section \ref{sec:PD}), corresponding to the refined signals ii)--iv), respectively. Each element indicates the augmented vector of observable channel parameters for the individual path, corresponding to the signals, defined as
\begin{align}
    \mathbf{z}_{k,j}^{\mathrm{D}} &\triangleq[\boldsymbol{\phi}_{k,j},\tau_{k,j},\boldsymbol{\theta}_{k,j}]^\top + \mathbf{r}_{k,j}^{\mathrm{D}},\\   
    \mathbf{z}_{k,j}^{\mathrm{O}} &\triangleq[\boldsymbol{\phi}_{k,j},\tau_{k,j},\boldsymbol{\theta}_{k,j}]^\top + \mathbf{r}_{k,j}^{\mathrm{O}},\\
    \mathbf{z}_{k,j}^{\mathrm{N}}&\triangleq[\bar{\tau}_{k,j},\boldsymbol{\theta}_{k,j}]^\top + \mathbf{r}_{k,j}^{\mathrm{N}},
\end{align}
    where $\mathbf{r}_{k,j}^{\mathrm{D}}$, $\mathbf{r}_{k,j}^{\mathrm{O}}$, and $\mathbf{r}_{k,j}^{\mathrm{N}}$ are respectively the Gaussian noises with the known covariance $\mathbf{R}^\mathrm{D}$, $\mathbf{R}^\mathrm{O}$, and $\mathbf{R}^\mathrm{N}$, which can be obtained by the \ac{fim} of the unknown channel parameters. 
    %
    %
    %
    We also consider false alarms caused by either the channel estimation error or detections of moving objects, only visible in a short time, modeled as clutter.
    The number of clutter components follows a Poisson distribution with  mean $\mu_\mathrm{Poi}^\mathrm{C}$.

\subsection{Merging of Double-bounce Measurements}

    We merge the double-bounce measurements $\mathcal{Z}^\mathrm{D}$ and $\mathcal{Z}^\mathrm{O}$ into a new set $\mathcal{Z}^\mathrm{R}$ by the ellipsoidal gating of two measurement sets~\cite{Panta2007_GatingPHD}. For each measurement $\mathbf{z}_{j}^\mathrm{D} \in \mathcal{Z}^\mathrm{D}$ and $\mathbf{z}_{j'}^\mathrm{O} \in \mathcal{Z}^\mathrm{O}$, we compute a distance metric
\begin{align}
    \mathrm{Dist}(j,j') =&\, 0.5[(\mathbf{z}_{j}^\mathrm{D}-\mathbf{z}_{j{'}}^\mathrm{O})^\top(\mathbf{R}_{j}^\mathrm{D})^{-1} (\mathbf{z}_{j}^\mathrm{D}-\mathbf{z}_{j{'}}^\mathrm{O})\nonumber \\
    &+(\mathbf{z}_{j}^\mathrm{D}-\mathbf{z}_{j{'}}^\mathrm{O})^\top(\mathbf{R}_{j'}^\mathrm{O})^{-1} (\mathbf{z}_{j}^\mathrm{D}-\mathbf{z}_{j{'}}^\mathrm{O})].
    \label{eq:EGD}
\end{align}
    If $\min_{j'}\mathrm{Dist}(j,j') < T_\mathrm{MG}$, $\mathbf{z}_{j}^\mathrm{D}$ and $\mathbf{z}_{j'}^\mathrm{O}$  are averaged and their average is added to $\mathcal{Z}^\mathrm{R}$ (with associated  covariance $\mathbf{R}^\mathrm{R} = 0.25(\mathbf{R}_j^\mathrm{D}+\mathbf{R}_{j'}^\mathrm{O})$).
    %
    Otherwise,  $\mathbf{z}_{j}^\mathrm{D}$ and  $\mathbf{z}_{j{'}}^\mathrm{O}$ are added to $ \mathcal{Z}^\mathrm{R}$.
    \subsection{Parallel PMB Filtering}
    
    We run two independent \ac{pmb} filters.
    One filter takes only the double-bounce measurements $\mathcal{Z}^\mathrm{R}$ as input   (by UE-RIS-SPs-UE and UE-SPs-RIS-UE signals), while  the other  takes the single-bounce measurements $\mathcal{Z}^\mathrm{N}$ as input (by UE-SPs-UE signal) for conventional \ac{nris}-sensing. 
    Each filter is a \ac{pmb} filter \cite{Hyowon_MPMB_TVT2022}, which treats both the measurements and SPs as random finite sets. While the implementation details are beyond the scope of this paper, 
    %
     %
    after the \ac{pmb} filtering, we have two \ac{pmb} posteriors, denoted by $f^\mathrm{R}(\mathcal{X})$ and $f^\mathrm{N}(\mathcal{X})$. The posteriors are parameterized by $\{\lambda^\mathrm{R}(\mathbf{x}),\{r^{\mathrm{R},i},f^{\mathrm{R},i}(\mathbf{x}) \}_{i\in \mathcal{I}^\mathrm{R}}\}$;  $\{\lambda^\mathrm{N}(\mathbf{x}),\{r^{\mathrm{N},i},f^{\mathrm{N},i}(\mathbf{x}) \}_{i\in \mathcal{I}^\mathrm{N}}\}$, where $r^i$ and $f^i(\mathbf{x})$ are respectively existence probability and spatial density for the $i$-th detected SP  and $\mathcal{I}$ denotes the number of detected SPs. The intensity function $\lambda({\mathbf{x}})=\mu f(\mathbf{x})$ and  spacial density $f^i({\mathbf{x}})$ are respectively represented by the uniform and Gaussian distributions.  
    The above components are computed by a nonlinear Kalman filtering~\cite{HaykinCKF2009}, and the mixture densities of \ac{pmb} are approximated to a single \ac{pmb} density by the marginalization of data association~\cite{Hyowon_MPMB_TVT2022}.


\subsection{Fusion of Two PMBs}
    
    We perform periodic fusion of the two 
    \ac{pmb} posteriors.\footnote{One can also run three parallel \ac{pmb} filters, one for UE-SP-UE measurements, one for UE-RIS-SP-UE measurements, and one for UE-SP-RIS-UE measurements. The proposed fusion can then be applied to any pair of \acp{pmb}. } Note that multiplication of the \acp{pmb} is not correct, as it will lead to double usage of measurements. 
    For the fusion, we adopt the \ac{gci} method~\cite{Batistelli_GCI_JSTSP2013} and
    fuse two \ac{pmb} posteriors $f^\mathrm{R}(\mathcal{X})$ and $f^\mathrm{N}(\mathcal{X})$ as follows
\begin{align}
    \bar{f}(\mathcal{X}) = \frac{f^\mathrm{R}(\mathcal{X})^{w^\mathrm{R}}f^\mathrm{N}(\mathcal{X})^{w^\mathrm{N}}}
    {\int f^\mathrm{R}(\mathcal{X}')^{w^\mathrm{R}}f^\mathrm{N}(\mathcal{X}')^{w^\mathrm{N}} \delta \mathcal{X}'},
    \label{eq:GCI}
\end{align}
    where $w^\mathrm{R}$ and $w^\mathrm{N}$ are the fusion weights such that $w^\mathrm{R}+w^\mathrm{N}=1$.
    The fused density $\bar{f}(\mathcal{X})$ is also a \ac{pmb}. 
    %
    Due to the variable detection probability and error variances, 
    the intensities and detected SP densities are separately fused, following the procedure in~\cite{Markus_PMBFusion_2020Access}. 

\begin{table}[t!]
\centering
\caption{Simulation parameters used in performance evaluation.}
\label{tab:SimulPara}
\begin{tabular}{ll}
	\hlineB{1}
	Parameter & Value\\ \hline 
	\ac{ris} array size & $N_\mathrm{R} = 2500$~($50\times50$) \\
	\ac{ue} array size & $N_\mathrm{U} = 16$~(4 by 4) \\
	No. transmissions & $T=40$ \\
	Carrier frequency & $f_c = 30$ GHz \\
	Speed of light & $c = 3 \times 10^8$ m/s \\
	Wavelength & $\lambda = 1$ cm \\
	Bandwidth &  $B=200$ MHz \\
	Subcarrier spacing & $\Delta_f = 120$ kHz \\
	No. subcarriers & $N_\mathrm{SC}= 1600$ \\
	Transmission power & $E_sN_\mathrm{SC} \Delta_f = 37$ dBm \\
	Noise variance & $ N_0 = -166$ dBm/Hz \\
	\hlineB{1}
\end{tabular}
\end{table}   

\begin{figure}[t!]
\begin{centering}
    \hspace{-1mm}
	{\input{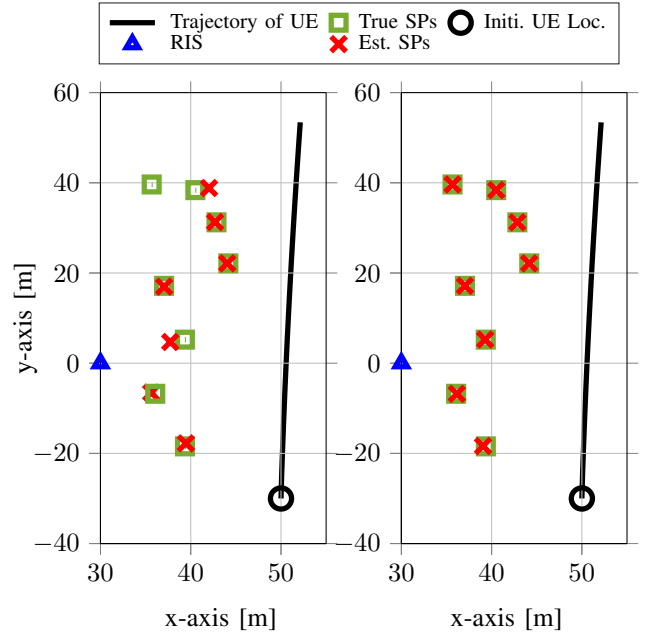}}\\
\vspace{-4mm}
	\caption{The \ac{ue} trajectory and map environment consisting of a single RIS and eight SPs for performance evaluation. The red x markers indicate the exemplary sensing results: (a) RIS-sensing, (b) NRIS-sensing, and (c) fusion of RIS- and NRIS sensing.}
	\label{Fig:TD_Map}
	\vspace{-5mm}
\end{centering}
\end{figure}

\section{Numerical Results}\label{sec:results}

\subsection{Simulation Setup}
    The simulation scenarios include a \ac{ue} moving along a predefined trajectory, a single \ac{ris}  attached on the wall, and eight \acp{sp}  distributed near the \ac{ue} trajectory, as shown in Fig.~\ref{Fig:TD_Map}. The  \ac{ris} location is set to $\mathbf{x}_\mathrm{R} = [30,0,20]^\top$, and \acp{sp} are randomly deployed in the space with the size of $(30,50)\times (-30,50)\times(2,10) $ m$^3$.
    The initial state is $\mathbf{s}_0 = [50,-30,0,\pi/2,11.11]^\top$, with units in meters for the first three, and radian and m/s for the latter two elements. 
    During $K=15$ time steps, the \ac{ue} dynamics follows the constant turn model~\cite{Li_Dynamic_TAES2003}, and the \ac{ue} states are known. 
    The simulation parameters used in performance evaluation are summarized in Table~\ref{tab:SimulPara}.
    
    We adopt random \ac{ris} phase profiles  ${\phi}_{\tilde{t},k}^n \sim \mathcal{U}[0,2\pi)$ for $n=1,\dots,N_\mathrm{R}$~\cite{Kamran_RISloc_ICC2022,Kamran_RISMobility_JSTSP2022}.
    We set $\alpha_{k,l}=\lvert \alpha_{k,l} \rvert e^{-j(2\pi f_c \tau_{k,l}+\nu_\mathrm{G})}$ and $\beta_{k,l}= \lvert \beta_{k,l} \rvert e^{-j(2\pi f_c \bar{\tau}_{k,l}+\nu_\mathrm{G})}$.
    For \ac{ris}~\cite{Zohair_5GFIM_TWC2018} and
    \ac{nris}~\cite{Wankai_RISModel_TWC2020} paths,
    the path gain amplitude models are adopted with $\lambda/4$ and $\lambda/2$ antenna spacing in \ac{ris}\footnote{Grating lobes at the \ac{ris} are avoided with  antenna spacing $d_\mathrm{R}\leq \lambda/4$~\cite{Kamran_RISloc_ICC2022}.} and \ac{ue}, respectively, given by
\begin{align}
    \lvert \alpha_{k,l} \rvert^2 &= \frac{E_s \lambda^2 (g_{\mathrm{UR},k})^{2q_0} }{16 (4\pi)^2 d_{\mathrm{UR},k}^2} 
\begin{cases}
   \frac{(g_{\mathrm{UR},k})^{2q_0}\lambda^2} {4\pi d_{\mathrm{UR},k}^2}, & l = 0\\
    \frac{(g_{\mathrm{SR}}^l)^{2q_0} \lambda^2 S_\mathrm{RCS}}
    {(4\pi)^2 (d_{\mathrm{SR}}^l)^2 (d_{\mathrm{SU},k}^l)^2},
        & l\neq 0
\end{cases}\notag\\
    \lvert \beta_{k,l} \rvert^2 &= \frac{E_s \lambda^2 S_\mathrm{RCS}}{(4\pi)^3 (d_{\mathrm{SU},k}^l)^4},~~~ l\neq 0,
    \notag
\end{align}
    where $\sqrt{E}_s$ is the energy per subcarrier,
    $\nu_\mathrm{G} \sim \mathcal{U}[0,2\pi)$ is the unknown phase offset, 
    $q_0 =0.285$, $g_{\mathrm{UR},k} = (\mathbf{x}_{\mathrm{U},k}-\mathbf{x}_{\mathrm{R}})^\top \mathbf{n}_\mathrm{R}/
    d_{\mathrm{UR},k}$, 
    $g_{\mathrm{SR}}^l = (\mathbf{x}^l-\mathbf{x}_{\mathrm{R}})^\top \mathbf{n}_\mathrm{R}/d_\mathrm{SR}^l$,
    $d_{\mathrm{UR},k}=\lVert \mathbf{x}_{\mathrm{U},k}-\mathbf{x}_{\mathrm{R}} \rVert$,
    $d_{\mathrm{SR}}^l=\lVert \mathbf{x}^l-\mathbf{x}_{\mathrm{R}} \rVert$,
    $d_{\mathrm{SU},k}^l=\lVert \mathbf{x}^l-\mathbf{x}_{\mathrm{U},k} \rVert$,
    $\mathbf{n}_\mathrm{R}=[1,0,0]^\top$  is the normal vector of the \ac{ris},
    and $S_\mathrm{RCS}= 50$~m$^2$ is the radar cross section.
    
    For the \ac{pmb} filter, we adaptively compute the \acp{dp} and utilize them in data association and measurement update~\cite{wymeersch2020adaptive}.
    In the update step,  $p_\mathrm{U}(\mathbf{x},\mathbf{s}_k) = 0.95$ for the intensity, and $p_\mathrm{U}^\mathrm{R}(\mathbf{x}_{k-1}^l,\mathbf{s}_k)=\max(p_\mathrm{U}^\mathrm{D}(\mathbf{x}_{k-1}^l,\mathbf{s}_k),p_\mathrm{O}^\mathrm{R}(\mathbf{x}_{k-1}^l,\mathbf{s}_k))$ and $p_\mathrm{U}^\mathrm{N}(\mathbf{x}_{k-1}^l,\mathbf{s}_k)$ for the Bernoulli densities.
    Here, $\mathbf{x}_{k-1}^l$ is the updated \ac{sp} location at the previous time step, and to compute the \ac{dp}, the transmission ratio is set to $T_1/T_2=1$ such that $T_1+T_2 = T/2$.
    The sensing performance are evaluated by the \ac{gospa}~\cite{RahmathullahGFS:2017}, averaged over 100 simulation runs.
    The visibility of individual path and measurement generation are determined by the proposed \acp{dp}, computed as~\eqref{eq:ADP} with $p_\mathrm{FA}=10^{-3}$.
    For the measurement noise covariance, we compute the \ac{fim} of the  channel parameters given the noiseless signals of \eqref{eq:NRIS}--\eqref{eq:reSRIS}.
    Other parameters for the \ac{pmb} filter follow \cite[Sec.~VI-A]{Hyowon_MPMB_TVT2022}.
    For the \ac{pmb} posterior fusion, we set the thresholds $T_\mathrm{MG} = 36$.

\begin{figure}[t!]
\begin{centering}
    \hspace{-1mm}\subfloat[\label{Fig:PDMap_U_randRIS}]
	{\includegraphics[width=0.45\columnwidth]
	{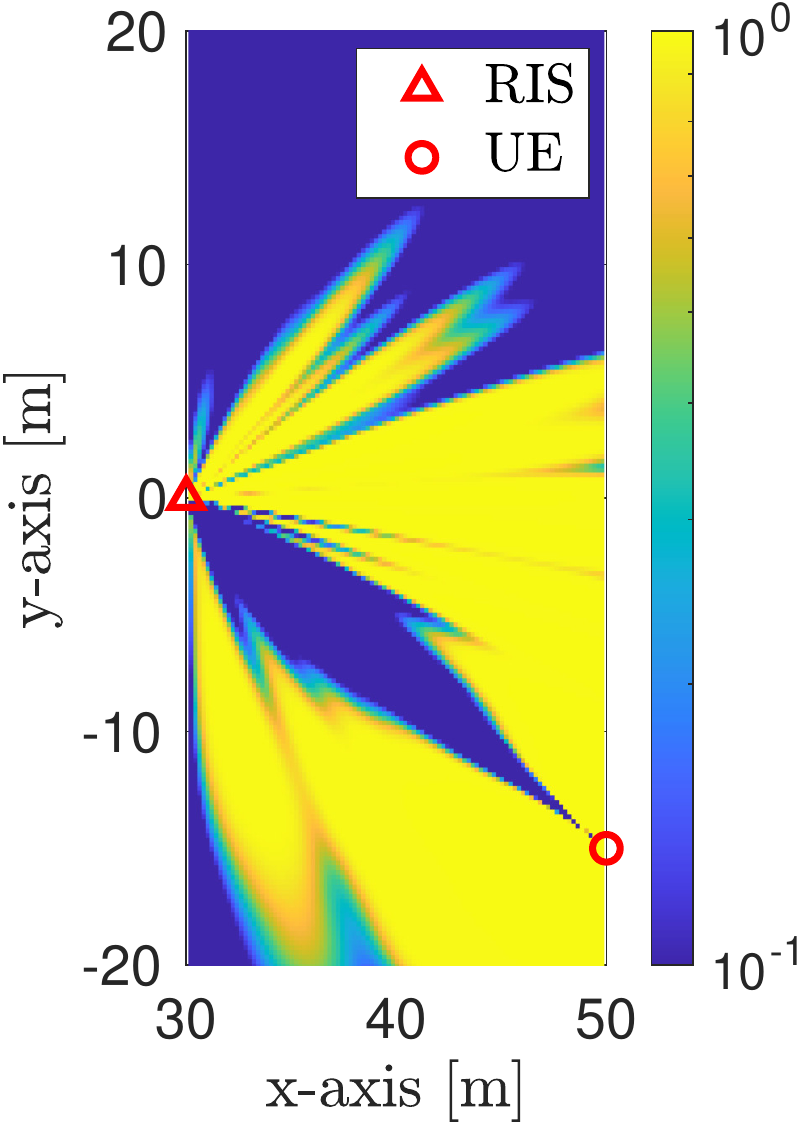}}
    \subfloat[\label{Fig:PDMap_S_randRIS}]
	{\includegraphics[width=0.45\columnwidth]{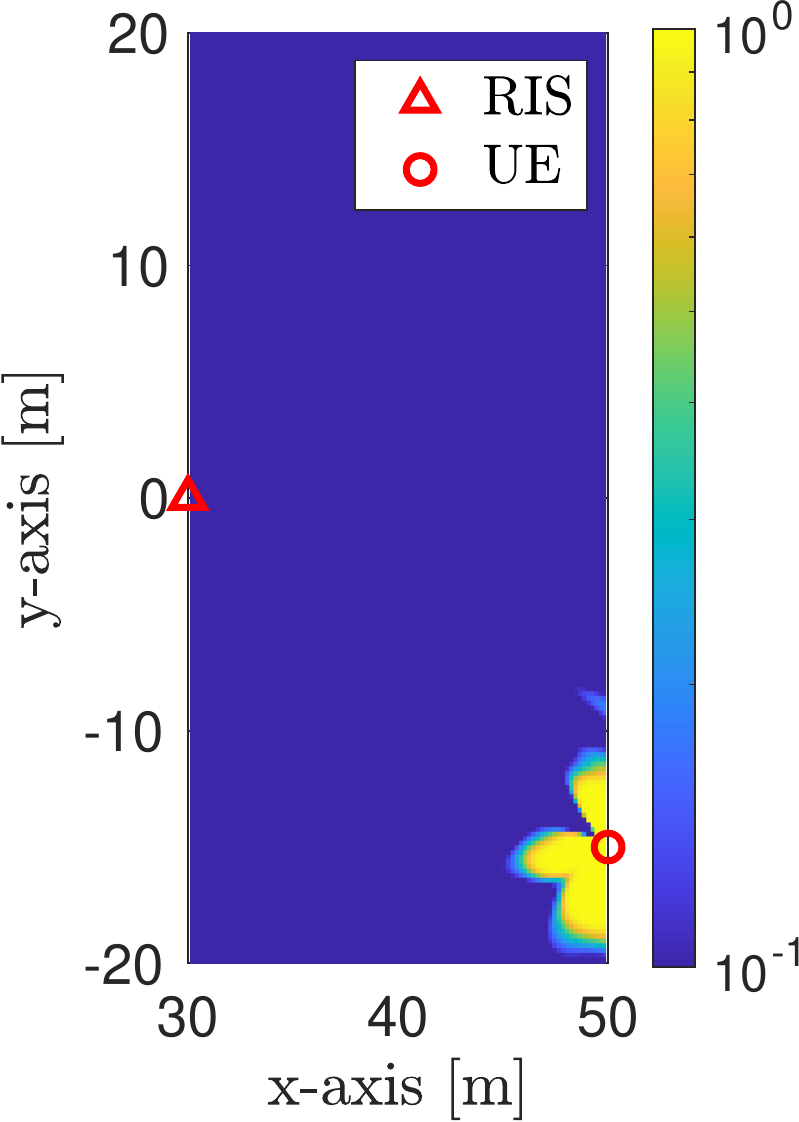}}
    \hspace{-1mm}\subfloat[\label{Fig:PDMap_U_directRIS}]
	{\includegraphics[width=0.45\columnwidth]
	{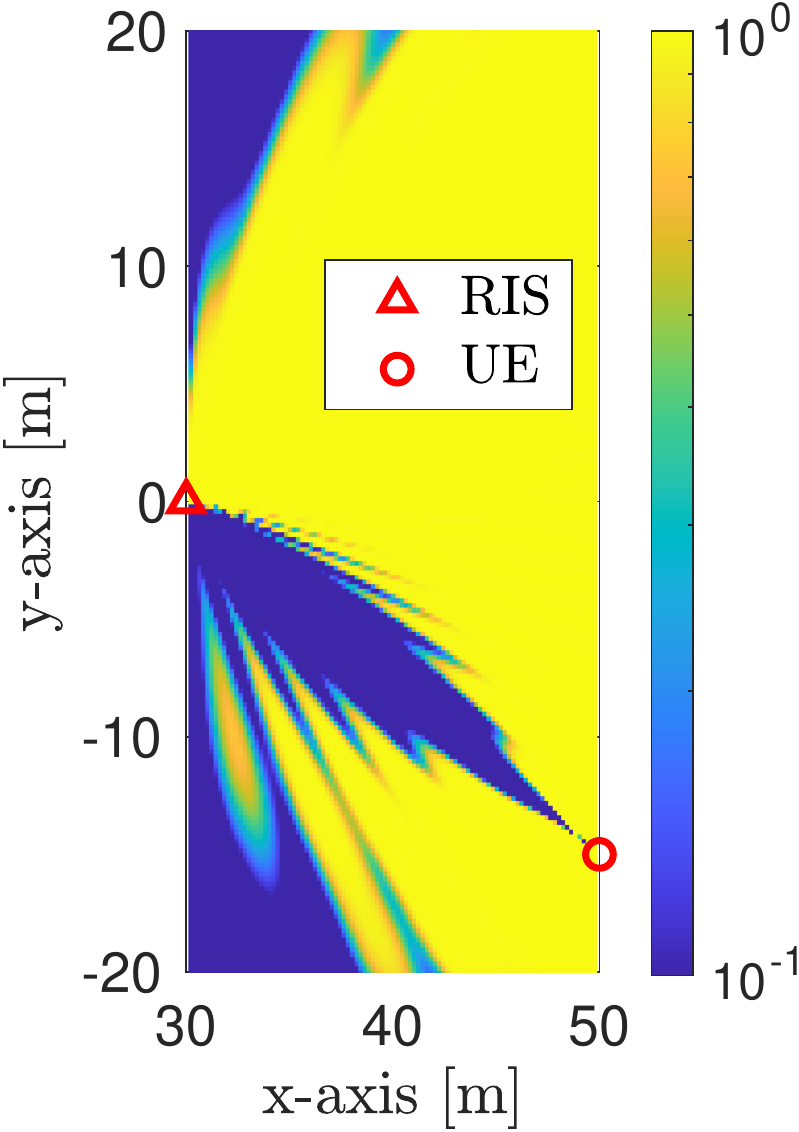}}
    \subfloat[\label{Fig:PDMap_S_directRIS}]
	{\includegraphics[width=0.45\columnwidth]{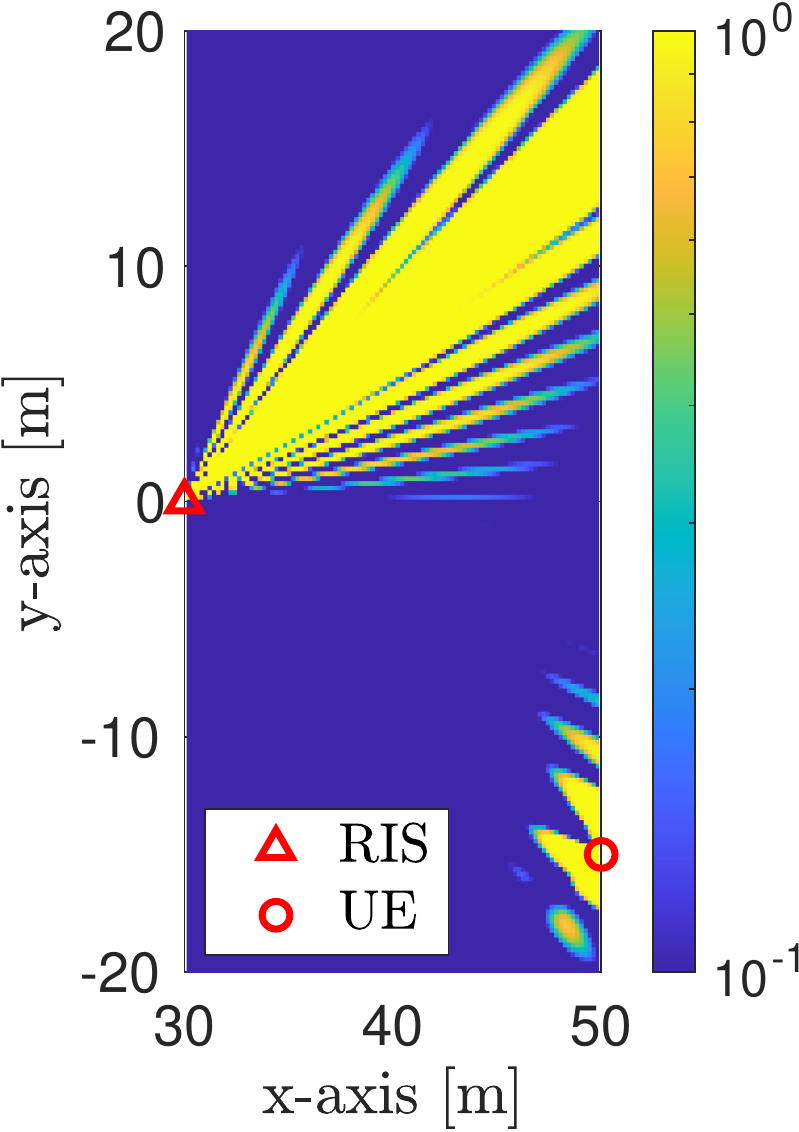}}
	\caption{Detection probabilities with the respect to the SP locations ($P_\mathrm{Tx}=20$ dBm, $T=20$, $\mathbf{x}_\mathrm{U}=[50,0,0]^\top$, $\mathbf{x}_\mathrm{R}=[30,0,0]^\top$): (a) UE-RIS-SP-UE with $\tilde{\bm{\omega}}_{\tilde{t}}^\mathrm{rand}$; (b) UE-SP-RIS-UE with $\tilde{\bm{\omega}}_{\tilde{t}}^\mathrm{rand}$, (c) UE-RIS-SP-UE with $\tilde{\bm{\omega}}_{\tilde{t}}^\mathrm{direct}$; and (d) UE-SP-RIS-UE with $\tilde{\bm{\omega}}_{\tilde{t}}^\mathrm{direct}$.
	We set $\tilde{\bm{\omega}}_{\tilde{t}}^\mathrm{direct} = (\mathbf{a}_\mathrm{U}(\bm{\phi}_{0}) \odot \mathbf{a}_\mathrm{U}(\bm{\phi}))^*$, where $\bm{\phi}$ is the angle at RIS to the point $[50,15,0]^\top$.
	}
	\label{Fig:PDMap}
 	\vspace{-.5cm}
\end{centering}
\end{figure}


\begin{figure}[t!]
\begin{centering}
    \vspace{4mm}
    \hspace{-1mm}
	{\input{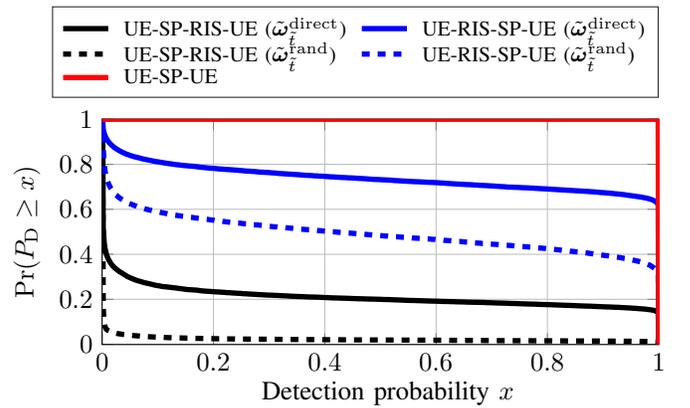}}
\vspace{-.6cm}
	\caption{Complementary cumulative distribution function of the detection probabilities $P_\mathrm{U}$ for UE-RIS-SP-UE, UE-SP-RIS-UE, UE-SP-UE paths with RIS configurations that are  random $\tilde{\bm{\omega}}_{\tilde{t}}^\mathrm{rand}$ or directional $\tilde{\bm{\omega}}_{\tilde{t}}^\mathrm{direct}$.
	}
	\label{Fig:PD_CDF}
 	\vspace{-.5cm}
\end{centering}
\end{figure}

\begin{figure}[t!]
\begin{centering}
    \hspace{-1mm}
	{
%
%
\definecolor{mycolor1}{rgb}{0.46667,0.67451,0.18824}%

\begin{tikzpicture}
\begin{axis}[%
width=74mm,
height=30mm,
at={(0mm,0mm)},
scale only axis,
xmin=0,
xmax=15,
ymin=0,
ymax=40,
axis background/.style={fill=white},
xmajorgrids,
ymajorgrids,
legend style={legend cell align=left, align=left, draw=white!15!black, style={row sep=-0.1cm}}
]
\addplot [color=black, line width=1.0pt, mark=square, mark options={solid, black}]
  table[row sep=crcr]{%
0	40\\
1	40\\
2	30.0359191755992\\
3	27.4110332567641\\
4	23.2463373603299\\
5	20.4664617190123\\
6	17.0259866973527\\
7	13.4763038757284\\
8	7.22551897011555\\
9	2.80461786556152\\
10	2.21699588010074\\
11	2.30790140140881\\
12	2.51686708792527\\
13	2.54174558325397\\
14	2.54174558325398\\
15	2.54174558326185\\
};
\addlegendentry{\footnotesize{RIS PMB}}

\addplot [color=blue, line width=1.0pt, mark=triangle, mark options={solid, blue}]
  table[row sep=crcr]{%
0	40\\
1	40\\
2	0.185488082203292\\
3	0.127403508547173\\
4	0.0852585713943821\\
5	0.0532559631673391\\
6	0.0305454733767466\\
7	0.018073476340395\\
8	0.0190880517316092\\
9	0.0283990940785252\\
10	0.0481657664228897\\
11	0.075997536684343\\
12	0.0989643146463669\\
13	0.113902872196415\\
14	0.16823211459857\\
15	0.241818960262488\\
};
\addlegendentry{\footnotesize NRIS PMB}

\addplot [color=red, dashed, line width=1.0pt, mark=x, mark options={solid, red}]
  table[row sep=crcr]{%
0	40\\
1	40\\
2	0.544991535990844\\
3	0.304532673133715\\
4	0.643473847031496\\
5	0.208090917872953\\
6	0.192811077223422\\
7	0.0486135530122147\\
8	0.0558474771340542\\
9	0.0711674430420473\\
10	0.0857621617434524\\
11	0.120309137926215\\
12	0.146202480278557\\
13	0.15784935811787\\
14	0.195680239746607\\
15	0.253290530024825\\
};
\addlegendentry{\footnotesize Fusion PMB}

\addplot [color=mycolor1, dashed, line width=1.0pt]
  table[row sep=crcr]{%
0	40\\
1	40\\
2	38.5518050385668\\
3	33.9561488253463\\
4	27.9005359290363\\
5	24.6364075610597\\
6	22.0618871744006\\
7	20.4679322236934\\
8	18.0494824973813\\
9	15.8745699833838\\
10	14.6631028610118\\
11	13.0958811137401\\
12	12.8614040590988\\
13	12.8614040602051\\
14	12.8614040642317\\
15	12.8614040653151\\
};
\addlegendentry{\footnotesize RIS PMB, random precoders}

\end{axis}

\node[rotate=0,fill=white] (BOC6) at (3.8cm,-.65cm){time $k$};
\node[rotate=90] at (-10mm,15mm){GOSPA dist. [m]};
\end{tikzpicture}%
}\vspace{-.4cm}
	\caption{GOSPA distances of SP sensing, with the directional and random precoders at the UE during the $T_1$ transmissions.
	}
	\label{Fig:Sensing}
 	\vspace{-.5cm}
\end{centering}
\end{figure}
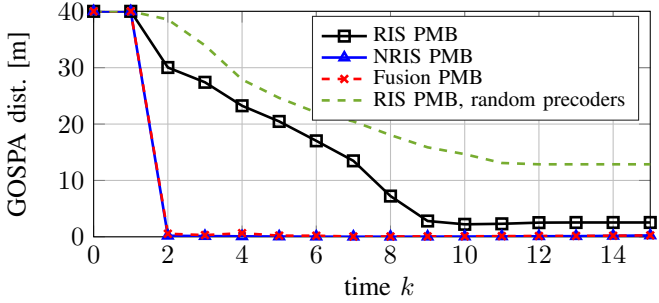


\subsection{Results and Discussions}

\subsubsection{Adaptive Detection Probability}
    Fig.~\ref{Fig:PDMap} depicts the \acp{dp} with the different \ac{sp} locations.
    If we use   $\tilde{\mathbf{f}}_{k,\tilde{t}}= \mathbf{a}^*_\mathrm{U}(\boldsymbol{\theta}_{k,0})/\Vert \mathbf{a}_\mathrm{U}(\boldsymbol{\theta}\Vert $ (i.e., directional precoder to the \ac{ris}), 
    there is a region between \ac{ris} to \ac{ue}, where \acp{sp} are not  detectable, as shown in Fig.~\ref{Fig:PDMap_U_randRIS}--Fig.~\ref{Fig:PDMap_U_directRIS}, where Fig.~\ref{Fig:PDMap_U_randRIS} applies random RIS configurations, while Fig.~\ref{Fig:PDMap_U_directRIS} applies  directional RIS configurations (i.e., $\tilde{\bm{\omega}}_{\tilde{t}}^\mathrm{direct} = (\mathbf{a}_\mathrm{U}(\bm{\phi}_{0}) \odot \mathbf{a}_\mathrm{U}(\bm{\phi}_{l}))^*$). 
    This occurs because the combiner $\mathbf{W}_\perp^\mathrm{H}$ is applied to extract the  UE-RIS-SP-UE signal and reject the UE-RIS-UE signal. In other area, the \ac{dp} is high, due to the high-gain directional UE precoding, and array gain due to the $N_{\mathrm{U}}-1$-dimensional observation. 
    %
    On the other hand, if we use a UE beam with null to the \ac{ris}, i.e., random precoders with  $\tilde{\mathbf{f}}_{k,\tilde{t}}^{\mathrm{H}} \mathbf{a}_\mathrm{U}(\boldsymbol{\theta}_{k,0})=0$, the resulting \ac{dp} 
is    shown in Fig.~\ref{Fig:PDMap_S_randRIS}--Fig.~\ref{Fig:PDMap_S_directRIS}.
    The UE-SP-RIS-UE path is illuminated with low gain random UE precoders and provides only a scalar observation after combining. 
    When using the directional RIS phase configurations,  
    the detectable region from RIS to the point point $[50,15,0]^\top$ is larger than that with random RIS configuration $\tilde{\bm{\omega}}_{\tilde{t}}^\mathrm{rand}$. We omit the \acp{dp} of the UE-SP-UE signals since they are approximately equal to 1.
    %
    %
    Fig.~\ref{Fig:PD_CDF} shows the \ac{ccdf}
    of \acp{dp} for all combinations of different \acp{sp} locations and \ac{ue} trajectories.
    We see that lower \acp{dp} are achieved in the double-bounce signals, compared to the UE-SP-UE signal, due to the severe path loss of \ac{ris} reflection.
    We achieved higher \acp{dp} in the UE-RIS-SP-UE, compared to the UE-SP-RIS-UE, due to  $N_{\mathrm{U}}-1$ beams by the combining matrix $\mathbf{W}_\perp \in\mathbb{R}^{N_{\mathrm{U}}-1 \times N_{\mathrm{U}}}$.
    Thanks to the directional RIS phase $\tilde{\bm{\omega}}_{\tilde{t}}^\mathrm{direct}$, we obtain higher \acp{dp} for the UE-RIS-SP-UE and UE-SP-RIS-UE signals. For a more in-depth discussion on the detection performance, we refer to Appendix \ref{app:LinkBudget}.

\subsubsection{Sensing}
    Fig.~\ref{Fig:Sensing} shows the sensing performance. The solid black curves with the square markers indicate the RIS-sensing performance by the \ac{pmb} filter given the measurement $\mathcal{Z}_k^\mathrm{D}$ and $\mathcal{Z}_k^\mathrm{O}$; solid blue curves with triangle markers indicate the NRIS-sensing performance given the measurement $\mathcal{Z}_k^\mathrm{N}$; and solid red curves with ‘x' markers indicate the performance of the fusion of RIS- and NRIS-sensing. 
    The \ac{pmb} posteriors for RIS- and NRIS-sensing are fused, merged into one map.
    %
    With the directional precoder $\tilde{\mathbf{f}}_{k,\tilde{t}}= \mathbf{a}^*_\mathrm{U}(\boldsymbol{\theta}_{k,0})/\Vert \mathbf{a}_\mathrm{U}(\boldsymbol{\theta}\Vert $ during the $T_1$ transmissions, the \ac{sp} \ac{gospa} distances gradually decrease as the number of observable \acp{sp} via double-bounce signals increases over time steps while the \acp{sp} via the UE-SP-UE are always observable. 
    In addition, the measurement noise covariances of the double-bounce signals are higher than the UE-SP-UE signal, due to the severe path loss.
    Therefore,
    the RIS-sensing performance is worse than the NRIS-sensing.  
    To show the importance of the directional UE precoders, we see that with random UE precoders during the $T_1$ transmissions, the \acp{sp} are rarely sensed via double-bounce signals, leading to poor GOSPA. 
\section{Conclusions}
    We presented a \ac{ris}-enabled passive object sensing framework  with a monostatic sensing UE and several passive objects. This problem is shown to be challenging due to the multiple observations of each objects, via both single- and double-bounce paths. 
    Detection probabilities for  different 
    paths and signals were derived and used in the observation system and \ac{pmb} filter.
    Using the expressions of the detection probabilities, we analyzed the impact of precoder and the \ac{ris}. Sensing methods for data association and fusion were proposed and evaluated. 
    Obtained results demonstrate that 
    double-bounce paths provided limited information in addition to single-bounce paths, 
    due to severe path loss of RIS reflection.
    
\vspace{-1mm}


\appendices
\vspace{-1mm}
\section{Channel Parameters}
\label{app:ChPara}
    We define the channel parameters as follows:
    $\tau_{k,0} = 2\lVert \mathbf{x}_{\mathrm{U},k}-\mathbf{x}_\mathrm{R} \rVert/c$, $\tau_{k,l} = (\lVert \mathbf{x}_{\mathrm{U},k}-\mathbf{x}_\mathrm{R} \rVert+\lVert \mathbf{x}^l-\mathbf{x}_\mathrm{R} \rVert + \lVert \mathbf{x}_{\mathrm{U},k}-\mathbf{x}^l \rVert)/c$,
    $\bar{\tau}_{k,l} =  2\lVert \mathbf{x}_{\mathrm{U},k}-\mathbf{x}^l \rVert/c$,
    $\phi_{k,0}^\mathrm{az} = \mathrm{atan2}(y_{\mathrm{UR}},x_{\mathrm{UR}})$, $\phi_{k,0}^\mathrm{el} = \mathrm{asin}(z_{\mathrm{UR}},\lVert \mathbf{x}_{\mathrm{UR}}\rVert)$,
    $\phi_{k,l}^\mathrm{az} = \mathrm{atan2}(y_{\mathrm{SR}}^l,x_{\mathrm{SR}}^l)$, $\phi_{k,l}^\mathrm{el} = \mathrm{asin}(z_{\mathrm{SR}}^l,\lVert \mathbf{x}_{\mathrm{SR}}^l\rVert)$,
    $\theta_{k,0}^\mathrm{az} = \mathrm{atan2}(y_{\mathrm{RU}},x_{\mathrm{RU}})$, 
    $\theta_{k,0}^\mathrm{el} = \mathrm{asin}(z_{\mathrm{RU}},\lVert \mathbf{x}_{\mathrm{RU}}\rVert)$,
    $\theta_{k,l}^\mathrm{az} = \mathrm{atan2}(y_{\mathrm{SU}}^l,x_{\mathrm{SU}}^l)$, $\phi_{k,l}^\mathrm{el} = \mathrm{asin}(z_{\mathrm{SU}}^l,\lVert \mathbf{x}_{\mathrm{SU}}^l\rVert)$,
    where $\mathbf{x}_{\mathrm{UR}} = \mathbf{O}_\mathrm{R}^\top(\mathbf{x}_{\mathrm{U},k} - \mathbf{x}_{\mathrm{R}})$, $\mathbf{x}_{\mathrm{SR}}^l = \mathbf{O}_\mathrm{R}^\top(\mathbf{x}^l - \mathbf{x}_{\mathrm{R}})$,
    $\mathbf{x}_{\mathrm{RU}} = \mathbf{O}_\mathrm{U}^\top(\mathbf{x}_{\mathrm{R}}-\mathbf{x}_{\mathrm{U},k})$, $\mathbf{x}_{\mathrm{SU}}^l = \mathbf{O}_\mathrm{U}^\top(\mathbf{x}^l - \mathbf{x}_{\mathrm{U},k})$. Here, $\mathbf{O}_{{\mathrm{D}}}^\top$ is the rotation matrix that rotates global to local coordinates at UE, and similarly for the $\mathbf{O}_{{\mathrm{R}}}^\top$ and the \ac{ris}~\cite{Zohair_5GFIM_TWC2018}.

\section{Link Budget Analysis}\label{app:LinkBudget}

    To understand the fundamental limits on the \ac{ris} signals for monostatic sensing, we investigate path losses for the different signals i)-iv), determined by the  received to transmitted power ratio. To this end, we consider a scenario where the UE and SP are towards the broadside of the RIS and focus only on the power, without accounting for beamforming or combining at the UE. We define $d_\mathrm{UR} = \lVert \mathbf{x}_{\mathrm{U}}-\mathbf{x}_\mathrm{R} \rVert$, $d_\mathrm{US}=\lVert \mathbf{x}_{\mathrm{U}}-\mathbf{x} \rVert$, and $d_\mathrm{RS} = \lVert \mathbf{x}_\mathrm{R}-\mathbf{x} \rVert$. With transmit power $P_\mathrm{T}$, 
    We will denote 
    the received powers for the different signals i)-iv) are respectively denoted by $P_\mathrm{R}^\mathrm{R}$, $P_\mathrm{R}^\mathrm{D}$, $P_\mathrm{R}^\mathrm{O}$, and $P_\mathrm{R}^\mathrm{N}$, given by
\begin{align}
    P_\mathrm{R}^\mathrm{R}
    & = 
    \frac{P_\mathrm{T} G_\mathrm{RIS}'\lambda^2 A_\mathrm{RIS} }{(4\pi)^2 d_\mathrm{UR}^4},\\
    P_\mathrm{R}^\mathrm{D} &= P_\mathrm{R}^\mathrm{O}
    = 
    \frac{P_\mathrm{T} G_\mathrm{RIS}\lambda^2 S_{\mathrm{RCS}} A_\mathrm{RIS}}{(4\pi)^4 d_\mathrm{US}^2d_\mathrm{UR}^2d_\mathrm{RS}^2},\\
    P_\mathrm{R}^\mathrm{N}
    &= 
    \frac{P_\mathrm{T}\lambda^2 S_{\mathrm{RCS}}}{(4\pi)^3 d_\mathrm{US}^4},
\end{align}
    where $A_{\mathrm{RIS}}=(\lambda/4)^2$ is the area of a RIS element,
    $G_{\mathrm{RIS}}=\mathbb{E}\left\{ |\mathbf{a}^{\top}(\bm{\phi}_{0})\bm{\Omega}\mathbf{a}(\bm{\phi}_{l})|^{2}\right\}$ is the RIS gain for the double-bounce paths, in which $\bm{\phi}_{l}$ is the \ac{aoa}/\ac{aod} from the \ac{sp}, and $G'_{\mathrm{RIS}}=\mathbb{E}\left\{ |\mathbf{a}^{\top}(\bm{\phi}_{0})\bm{\Omega}\mathbf{a}(\bm{\phi}_{0})|^{2}\right\}$ is the RIS gain for the UE-RIS-UE path. In this setup, $\bm{\phi}_{l}=\bm{\phi}_{0}$, so $G_{\mathrm{RIS}}=G'_{\mathrm{RIS}}$. In the case of directional RIS configurations $G_{\mathrm{RIS}}=N_{\mathrm{R}}^{2}$, while for random configurations 
    $G_{\mathrm{RIS}}=N_{\mathrm{R}}$. 
    
    We now plot the path loss $P_\mathrm{R}/P_\mathrm{T}$ for each of the paths in Fig.~\ref{Fig:LB}. We consider 2 scenarios: in scenario (a) the SP is between the UE and the RIS, so that $d_\mathrm{RS} = \rho_\mathrm{S}d_\mathrm{UR}$, for $\rho_\mathrm{S}\in(0,1)$; in scenario (b) the UE is between the SP and the RIS, so that $d_\mathrm{RU} = \rho_\mathrm{U}d_\mathrm{RS}$, for $\rho_\mathrm{U}\in(0,1)$. We set $d_\mathrm{UR}=30~\mathrm{m}$ for scenario (a) and $d_\mathrm{RS}=30~\mathrm{m}$ for scenario (b), while other parameters are as in Section \ref{sec:results}. 
    We observe that in scenario (a) (see Fig.~\ref{Fig:LB_Ex1}), the single bounce path UE-SP-UE is nearly always the strongest (with path loss above $-114$ dB). Under random configurations, the UE-RIS-UE path has a loss of around $-150$ dB, while the double-bounce path has a loss that varies from  $-130$ dB (SP close to RIS or UE) to $-160$ dB (SP in the middle). With directional RIS profiles all RIS paths are boosted by $10 \log_{10}(2,500)=34$ dB, providing a gain over the UE-SP-UE path with about $16$ dB when the SP is very close to the RIS. Moreover, in both cases of RIS configurations, the path UE-RIS-UE is generally stronger than the double-bounce paths, leading to severe interference (which was mitigated in this work by UE beamforming and combining).  In scenario (b), the curves for UE-SP-UE and UE-RIS-SP-UE are the same as in scenario (a), due to the symmetry of the path loss. The difference lies in the UE-RIS-UE path, which is stronger when the UE is close to the RIS, but again nearly always dominates and thus interferes with the double-bounce paths.

\begin{figure}[t!]
\begin{centering}
    \vspace{-2mm}
    \hspace{-1mm}
    \subfloat[SP between RIS and UE\label{Fig:LB_Ex1}]{\input{Figures/LB_Ex1.tex}}
    \\
    \vspace{-3mm}
	\subfloat[UE between RIS and SP\label{Fig:LB_Ex2}]
	{\input{Figures/LB_Ex2.tex}}
	\caption{Path losses where in (a) $d_\mathrm{RS} = \rho_\mathrm{S}d_\mathrm{UR}$ while in (b) $d_\mathrm{RU} = \rho_\mathrm{U}d_\mathrm{RS}$. 
	}
	\label{Fig:LB}
 	\vspace{-.5cm}
\end{centering}
\end{figure}

\balance
\bibliographystyle{IEEEtran}
\bibliography{IEEEabrv,bibliography}

\end{document}